\documentstyle[12pt,graphicx,amsfonts]{article}
\begin{document}
\baselineskip 0.6cm

\def\simgt{\mathrel{\lower2.5pt\vbox{\lineskip=0pt\baselineskip=0pt
           \hbox{$>$}\hbox{$\sim$}}}}
\def\simlt{\mathrel{\lower2.5pt\vbox{\lineskip=0pt\baselineskip=0pt
           \hbox{$<$}\hbox{$\sim$}}}}
\def\x{x}

\begin{titlepage}

\begin{flushright}
UFIFT-HEP-04-4 \\
CLNS 04/1881 \\
UCB-PTH 04/17 \\
LBNL-55247
\end{flushright}

\vskip 1.2cm

\begin{center}

{\Large \bf 
Relaxing the Upper Bound on the Mass of the Lightest Supersymmetric Higgs Boson
}

\vskip 0.9cm

{\large
Andreas Birkedal$^{a}$, Z. Chacko$^{b}$ and Yasunori Nomura$^{c,d}$
}

\vskip 0.5cm

$^a$ {\it Department of Physics, University of Florida,
                Gainesville, FL 32611} \\
$^b$ {\it Department of Physics, University of Arizona,
                Tucson, AZ 85721} \\
$^c$ {\it Department of Physics, University of California,
                Berkeley, CA 94720} \\
$^d$ {\it Theoretical Physics Group, Lawrence Berkeley National Laboratory,
                Berkeley, CA 94720} \\

\vskip 1.2cm

\abstract{
We present a class of supersymmetric models in which the lightest Higgs-boson 
mass can be as large as a few hundred GeV ($200\!\sim\!300~{\rm GeV}$) while 
the successful MSSM prediction for gauge coupling unification is preserved. 
The theories are formulated on a 5D warped space truncated by two branes, 
and a part of the Higgs sector is localized on the infrared brane.  The 
structure of the Higgs sector in the four dimensional effective theory 
below the Kaluza-Klein mass scale is essentially that of the next-to-minimal 
supersymmetric standard model (NMSSM), or related theories. However, large 
values of the NMSSM couplings at the weak scale are now possible as these 
couplings are required to be perturbative only up to the infrared cutoff 
scale, which can in general be much lower than the unification scale. 
This allows the possibility of generating a large quartic coupling in the 
Higgs potential, and thereby significantly raising the Higgs-boson mass bound. 
We present two particularly simple models.  In the first model, the quark 
and lepton fields are localized on the ultraviolet brane, where the grand 
unified symmetry is broken.  In the second model, the quark and lepton fields 
are localized on the infrared brane, and the unified symmetry is broken both 
on the ultraviolet and infrared branes.  Our theories potentially allow the 
possibility of a significant reduction in the fine-tuning needed for correct 
electroweak symmetry breaking, although this is somewhat model dependent. 
}

\end{center}
\end{titlepage}

\section{Introduction}

Postulating supersymmetry as a solution to the hierarchy problem results 
in significant constraints on the mass of the lightest Higgs boson. 
For instance, in the minimal supersymmetric standard model (MSSM), the 
theoretical upper limit on the lightest neutral Higgs-boson mass is known 
to be around $130~{\rm GeV}$, assuming all superpartners have masses below 
a ${\rm TeV}$~\cite{Okada:1990vk,Carena:1995wu}.  This is also, roughly, 
the bound that arises in the minimal extension of the MSSM, so called the 
next-to-minimal supersymmetric standard model (NMSSM)~\cite{Nilles:1982dy}, 
assuming that the NMSSM singlet couplings remain perturbative up to the 
scale of gauge coupling unification, $\Lambda_{\rm GUT}$~\cite{Kane:1992kq}. 
This upper limit is complemented on the lower side by experimental 
data constraining the lightest Higgs-boson mass to be above $114~{\rm GeV}$ 
for many interesting regions of parameter space~\cite{unknown:2001xx}. 
Thus, it becomes important to understand the theoretical implications for 
supersymmetry if the lightest neutral Higgs boson is discovered to have a 
mass significantly above $130~{\rm GeV}$.

In this paper we present a class of supersymmetric models in which the 
lightest Higgs-boson mass can be as large as about a few hundred GeV 
($200\!\sim\!300~{\rm GeV}$) while preserving the successful MSSM prediction 
for gauge coupling unification.  Recall that in the NMSSM the existing tight 
upper limit for the lightest Higgs mass arises from the requirement that 
the couplings for the Higgs fields are all perturbative up to the unification 
scale, $\Lambda_{\rm GUT}$.  However, this requirement does not necessarily 
have to be imposed if, for example, the Higgs fields are composite (or 
mixtures of elementary and composite fields) arising at low energies. 
In this case the couplings of the singlet field in the NMSSM, for example, 
can become non-perturbative at a scale much below $\Lambda_{\rm GUT}$, 
significantly weakening the upper bound of the lightest Higgs-boson mass. 
The crucial point is that the low-energy non-perturbative dynamics in 
the Higgs sector does not necessarily mean that the entire theory enters 
into a non-perturbative regime below $\Lambda_{\rm GUT}$.  In fact, it is 
perfectly possible that the MSSM quark, lepton and gauge sectors stay 
perturbative up to the scale $\Lambda_{\rm GUT}$, allowing a perturbative 
treatment for gauge coupling evolution, even if the Higgs sector becomes 
strongly interacting at low energies.  Then, as long as the strong dynamics 
in the Higgs sector respect an approximate global $SU(5)$ symmetry at 
energies above the TeV scale, the predicted values for the low-energy 
gauge couplings are the same as those in the MSSM. 

How do we explicitly realize the scenario described above?  An attractive 
way of dealing with strong dynamics is to consider higher dimensional 
theories.  Through the AdS/CFT duality~\cite{Maldacena:1997re}, as applied 
to the truncated AdS space~\cite{Arkani-Hamed:2000ds}, the above scenario 
is related to supersymmetric theories in five-dimensional (5D) warped space 
bounded by two branes~\cite{Randall:1999ee}.  The strongly interacting 
Higgs sector, then, corresponds to the Higgs (and singlet) fields localized 
to the infrared brane, or the TeV brane, while the perturbative sector 
corresponds to quarks and leptons (and a part of the Higgs) localized towards 
the ultraviolet brane, or the Planck brane, with the standard model gauge 
fields propagating in the bulk.  This setup is very attractive because 
supersymmetry can be broken on the infrared brane, allowing us to naturally 
understand the hierarchically small supersymmetry-breaking scale through 
the AdS warp factor~[\ref{Gherghetta:2000qt:X}~--~\ref{Nomura:2004it:X}] 
(see also~[\ref{Gherghetta:2000kr:X}~--~\ref{Chacko:2004mi:X}]). 
As shown in~\cite{Goldberger:2002pc}, this class of theories leaves 
many of the most attractive features of conventional unification intact; 
in particular, the successful MSSM prediction for gauge coupling 
unification is preserved, provided that the 5D bulk possesses an $SU(5)$ 
gauge symmetry which is broken at the Planck brane and that matter 
and two Higgs doublets are localized towards the Planck brane or 
have conformally-flat wavefunctions. (The successful prediction was 
anticipated earlier in~\cite{Pomarol:2000hp}, and techniques for 
calculating gauge coupling evolution in warped space were developed 
in~[\ref{Randall:2001gc:X}~--~\ref{Randall:2002qr:X}].)  These 
theories also have several nice features if the $SU(5)$ breaking at the 
Planck brane is caused by boundary conditions.  The models we present 
in this paper preserve the aforementioned attractive features, including 
the MSSM prediction for gauge coupling unification.  Alternative approaches 
to raising the upper bound on the mass of the lightest supersymmetric Higgs 
boson have been proposed recently in~\cite{Batra:2003nj,Casas:2003jx,%
Harnik:2003rs,Chang:2004db}.  Earlier work on raising the Higgs 
mass bound can be found, for example, in Refs.~\cite{Cvetic:1997ky,%
Casas:2001xv}.  We will comment on the relation of some of these 
papers to our work in later sections.

A large value for the physical Higgs-boson mass has the virtue that it 
potentially reduces the fine-tuning needed to break the electroweak symmetry 
at the correct scale in supersymmetric models.  It is known that in the MSSM 
we need a relatively large top-squark mass of $m_{\tilde{t}} \simgt 
500~{\rm GeV}$ in order to obtain an experimentally allowed physical 
Higgs-boson mass of $m_{\rm Higgs} \simgt 114~{\rm GeV}$ in generic 
parameter regions.  This large top-squark mass then leads to a large soft 
Higgs mass-squared parameter through radiative corrections given by 
$m_h^2 \simeq -(3y_t^2/4\pi^2)\, m_{\tilde{t}}^2 \ln(M_{\rm mess}/m_{\tilde{t}})$, 
where $M_{\rm mess}$ is the scale at which supersymmetry breaking is 
mediated to the MSSM sector.  This generically leads to fine-tuning 
of electroweak symmetry breaking as $m_h^2$ is typically larger than 
$m_{\rm Higgs}^2$ by an order of magnitude or larger, especially when 
$\ln(M_{\rm mess}/m_{\tilde{t}})$ is large.  Because $m_{\rm Higgs}$ 
can be as large as $200\!\sim\!300~{\rm GeV}$ in our theory, this 
tuning could potentially be reduced by a large amount.  Moreover, 
in warped supersymmetric models $\ln(M_{\rm mess}/m_{\tilde{t}})$ 
is generically small (see~\cite{Nomura:2003qb}),%
\footnote{For alternative ideas to reduce the fine-tuning, see 
e.g.~\cite{Kobayashi:2004pu,Birkedal:2004xi,Chankowski:2004mq}.}
so that our theory allows relatively large superpartner masses for 
a given value of $m_h^2$, which can be as large as $m_h^2 \simeq 
m_{\rm Higgs}^2/2 \simeq (150\!\sim\!200~{\rm GeV})^2$ without any 
fine-tuning.  In some of the explicit realizations of our theory, the virtue 
of these properties is somewhat reduced by the fact that there can be 
large tree-level Higgs soft masses, but we believe it is significant that 
we can construct simple models accommodating these features. 

The organization of the paper is as follows.  In the next section we 
analyze the NMSSM Higgs sector with the cutoff lowered to $\Lambda \ll 
\Lambda_{\rm GUT}$ and see that it can push up the bound on the 
lightest Higgs-boson mass to be as large as about $300~{\rm GeV}$. 
In section~\ref{sec:model-1} we construct a theory allowing a lowered cutoff 
for the Higgs sector while preserving the successful MSSM prediction for 
gauge coupling unification. The theory is formulated in a 5D supersymmetric 
warped space, and we also briefly discuss phenomenological consequences of the 
theory.  We can obtain the mass of the lightest Higgs boson as large as about 
$200~{\rm GeV}$ in this theory.  In section~\ref{sec:model-2} we present 
an alternative model possessing similar properties but having a different 
configuration of fields in the warped extra dimension.  This model allows 
the lightest Higgs-boson mass as large as $300~{\rm GeV}$, although the 
model requires an imposition of additional symmetries to be fully realistic. 
Conclusions are given in section~\ref{sec:concl}.  Some preliminary 
results of this paper were presented by one of the authors in~\cite{AB}.

\section{Lowering the Cutoff of the NMSSM Higgs Sector}
\label{sec:NMSSM}

The tree-level Higgs-boson mass bound in the MSSM is limited by the 
$Z$-boson mass $m_Z$.  However, the one-loop corrections from top quarks 
and squarks are sizable, leading to the upper limit of about $130~{\rm GeV}$. 
This situation is ameliorated only slightly by going to the NMSSM, which 
includes an extra gauge singlet $S$ that couples to the Higgs fields 
through the superpotential:
\begin{equation}
  W_{\rm NMSSM} 
    = \lambda S H_u H_d - \frac{\kappa}{3} S^3 
      + {\rm Yukawa\,\, couplings}.
\label{eq:superpot}
\end{equation}
Including the singlet couplings in the one-loop Higgs mass bound results in 
\begin{equation}
  m_{h,1-{\rm loop}}^2 
    \leq m_{Z}^2 \cos^2\!2\beta + \lambda^2 v^2 \sin^2\!2\beta 
      + \frac{3}{4\pi^2} y_t^4 v^2 \sin^4\!\beta\, 
        \ln\Bigl(\frac{m_{\tilde{t}_1}m_{\tilde{t}_2}}{m_t^2}\Bigr),
\label{eq:1loopbound}
\end{equation}
where $m_{\tilde{t}_1}$ and $m_{\tilde{t}_2}$ are the masses of the two 
top squarks, $m_t$ is the top-quark mass, $y_t$ is the top Yukawa coupling, 
$\tan\beta \equiv \langle H_u \rangle/\langle H_d \rangle$, and $v \equiv 
(\langle H_u \rangle^2 + \langle H_d \rangle^2)^{1/2}$.  Here we have 
set the mixing between the left- and right-handed top squarks to be zero 
for simplicity, and neglected a one-loop correction arising from the 
coupling $\lambda$.  At first sight, Eq.~(\ref{eq:1loopbound}) appears 
to be a significant relaxation of the upper bound on the Higgs-boson mass. 
However, the value of $\lambda$ at the weak scale is generally suppressed 
if we require $\lambda$ and $\kappa$ to remain perturbative up to the 
scale of gauge coupling unification, $\Lambda_{\rm GUT} \simeq 2 \times 
10^{16}~{\rm GeV}$.  Once this requirement is imposed, one finds that 
the inclusion of the singlet field and its coupling to the Higgs fields 
increases the upper limit on the lightest Higgs mass by only about 
$10~{\rm GeV}$ compared with the MSSM~\cite{Ellwanger:1993hn}.

\begin{figure}[t]
\begin{center}
  \includegraphics[scale=0.6]{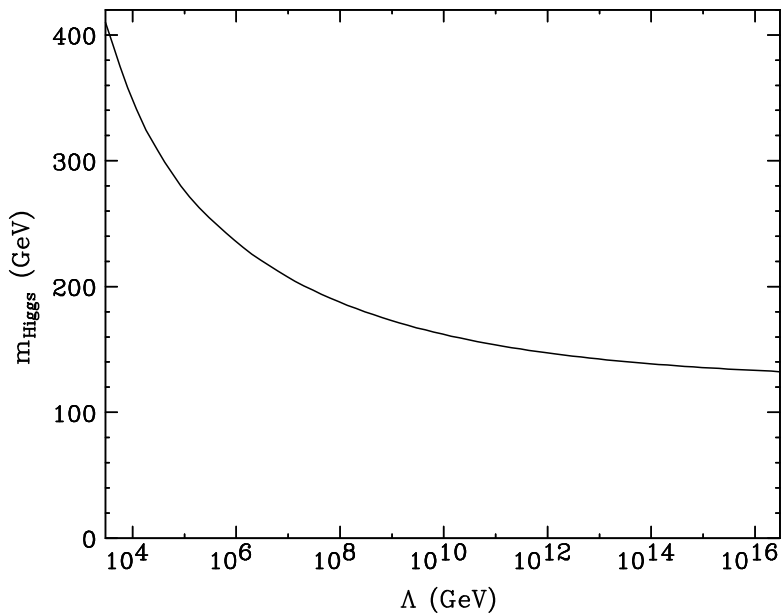}
\caption{Theoretical upper limits on the lightest Higgs-boson mass as 
 a function of the cutoff scale $\Lambda$ of the NMSSM Higgs sector.}
\label{fig:limplot}
\end{center}
\end{figure}

Now we ask what happens if we do not require perturbativity of the 
couplings in Eq.~(\ref{eq:superpot}) up to the unification scale.  One 
can see from the renormalization group equations (RGEs) for $\lambda$ 
and $\kappa$ that the value of $\lambda$ always decreases when it is 
run down from a high scale:
\begin{eqnarray}
  \frac{d \lambda}{d\ln\mu} 
    &=& \frac{\lambda}{16 \pi^2} \left( 4 \lambda^2 + 2 \kappa^2 
      + 3 y_t^2 + 3 y_b^2 + y_\tau^2 - 3 g^2 - g'^2 \right),
\label{eq:RGElam} \\
  \frac{d \kappa}{d\ln\mu} 
    &=& \frac{6\kappa}{16 \pi^2} \left( \lambda^2 + \kappa^2 \right),
\label{eq:RGEk}
\end{eqnarray}
where $y_b$ and $y_\tau$ are the bottom and tau Yukawa couplings, and 
$g$ and $g'$ are the $SU(2)_L$ and $U(1)_Y$ gauge couplings. For instance, 
if we assume that $\lambda$ becomes non-perturbative at a scale $\Lambda$ 
above the weak scale but below $\Lambda_{\rm GUT}$, i.e. $\lambda(\Lambda) 
= 2\pi$, then a lower value of $\Lambda$ results in a higher value of 
$\lambda$ when it is run down to the weak scale. This, therefore, results 
in a higher upper limit on the mass of the lightest Higgs boson.  We 
have illustrated this in Fig.~\ref{fig:limplot}, where we have taken 
$\kappa(\Lambda)=0$ and $\lambda(\Lambda)= 2\pi$, which maximizes the 
Higgs-boson mass, and chosen the value of $\tan\beta$ such that the 
largest Higgs mass is obtained for each value of $\Lambda$.  A similar
analysis was performed earlier in~\cite{Tobe:2002zj}.
The improvement gained by lowering the scale at which the Higgs-singlet 
sector becomes non-perturbative is clear.  A scale of $\Lambda = 
10^4~{\rm GeV}$, for example, results in an upper Higgs mass bound 
of approximately $340~{\rm GeV}$, and lower values of $\Lambda$ give 
even larger Higgs-boson masses.  Although there are many uncertainties 
in the Higgs mass values obtained in this way, for example those arising 
from effects of non-zero values for $\kappa$, one-loop effects involving 
$\lambda$, higher-order effects and so on, we expect that we can still 
obtain the lightest Higgs-boson mass as large as about $300~{\rm GeV}$, 
especially for smaller values of $\Lambda$.  Note that this result applies 
to more general superpotentials of the form
\begin{equation}
  W_{\rm NMSSM} 
    = \lambda S H_u H_d + f(S)
      + {\rm Yukawa\,\, couplings},
\label{eq:superpot-2}
\end{equation}
where $f(S)$ is a general function of $S$: $f(S) = - \Lambda_S^2 S 
- (M_S/2) S^2 - (\kappa/3) S^3 + \cdots$.  Here $\Lambda_S$ and $M_S$ 
are mass parameters of order the weak scale.  The NMSSM is clearly 
a special case of this. 

In deriving the bound on the Higgs-boson mass in Fig.~\ref{fig:limplot}, 
we required the top Yukawa coupling to be perturbative ($y_t \simlt \pi$) 
only up to the scale $\Lambda$ and did not require this coupling to be 
perturbative up to the unification scale of $\Lambda_{\rm GUT} \simeq 
2 \times 10^{16}~{\rm GeV}$.  However, we have required all the other 
sectors of the theory, in particular the gauge couplings, to remain 
perturbative up to the unification scale.  Our next task is to construct 
explicit models that realize this type of behavior while preserving 
the successful MSSM prediction for gauge coupling unification.  In the 
next section, we present a model that shares certain features with the 
low cutoff NMSSM discussed here, although the resulting upper limits on 
the Higgs-boson mass are tighter than the naive values obtained here. 
A model that actually realizes the upper limits as high as those given in 
Fig.~\ref{fig:limplot} will be constructed in section~\ref{sec:model-2}.

\section{Supersymmetric Theory with a Heavy Higgs Boson}
\label{sec:model-1}

\subsection{Basic scheme}

In this section we construct a supersymmetric model allowing a relatively 
heavy (two hundred GeV or so) Higgs boson while preserving the successful 
MSSM gauge coupling prediction.  We have just seen that a simple way around 
the tight upper bound on the lightest Higgs-boson mass is to allow the Higgs 
sector to become non-perturbative at a scale below the gauge unification 
scale, $\Lambda_{\rm GUT}$.  To preserve the perturbative prediction for 
gauge coupling unification, this must be done in a way such that the sector 
relevant for the gauge coupling prediction remains perturbative up to 
$\Lambda_{\rm GUT}$.  As outlined in the introduction, this can be done 
by formulating the theory in 5D warped space truncated by two branes.

In a warped extra dimension, the effective cutoff scale of the theory 
changes with position in the fifth coordinate.  We denote this extra 
dimension by $y$, where $0 \leq y \leq \pi R$.  This can be thought of as 
arising from compactification on the orbifold $S^1/Z_2$.  Two branes exist 
in this setup: an infrared (IR) brane at $y=\pi R$ and an ultraviolet (UV) 
brane at $y=0$.  The brane tension causes a warping of the extra dimension 
into a slice of AdS space.  The AdS space describing the extra dimension 
is defined by the metric: 
\begin{equation}
  ds^2 = e^{-2k|y|} \eta_{\mu\nu} dx^\mu dx^\nu + dy^2,
\label{eq:metric}
\end{equation}
where $k$ denotes the curvature of the AdS space.  The four dimensional (4D) 
Planck scale $M_4$ is then related to the 5D Planck scale $M_5$ by $M_4^2 
\simeq M_5^3/2k$ (for $\pi kR \simgt 1$).  Here we take $k \sim M_5 \sim 
M_4$, but with $k$ a factor of a few smaller than $M_5/\pi$ so that the 
5D theory is under theoretical control.  Identifying $M_5$ with the cutoff 
of the 5D theory, the effective $y$-dependent cutoff, i.e. the cutoff 
scale measured in terms of the 4D metric $\eta_{\mu\nu}$, is given by 
\begin{equation}
  \Lambda_{\rm cutoff} = M_5 e^{-k|y|}. 
\label{eq:eff-cutoff}
\end{equation}
In particular, the effective cutoff on the IR brane is just $\Lambda_{\rm IR} 
= M_5 e^{-\pi kR}$, while the cutoff on the UV brane is just the 5D Planck 
scale, $\Lambda_{\rm UV} = M_5$.  The characteristic scale for the Kaluza-Klein 
(KK) excitations, which we call the compactification scale, is given 
by $M_c = \pi k\, e^{-\pi kR}$.  Since we assume $M_5 > \pi k$, the IR 
cutoff scale is larger than the masses of the first few KK states: 
$\Lambda_{\rm IR} > M_c$.  

Since the warped extra dimension gives a different effective cutoff at each point 
in $y$, fields located in different positions in the extra dimension see different 
cutoff scales.  Since we need our Higgs sector to have a low cutoff scale, we 
place a part of the Higgs sector, consisting of three chiral superfields $S$, 
$H^{(b)}_u$ and $H^{(b)}_d$ and the superpotential of Eq.~(\ref{eq:superpot-2}) 
with $H_u$ and $H_d$ replaced by $H^{(b)}_u$ and $H^{(b)}_d$, on the IR brane 
(the reason for the superscripts on the Higgs fields will become clear later). 
The cutoff for the Higgs sector is then given by $\Lambda_{\rm IR}$, which 
is thus identified with the $\Lambda$ of the previous section.  The matter 
fields are localized on the UV brane so that the cutoff for these fields is 
around the 4D Planck scale. This suppresses potentially dangerous proton 
decay and flavor-changing processes.  An important point is that, although 
the Higgs sector becomes non-perturbative at a low energy scale of 
$\Lambda_{\rm IR}$, it does not affect the physics of the rest of the model. 
In particular, the evolution of the gauge couplings is not affected by this 
non-perturbativity of the Higgs sector.  This is because physics above the 
scale $M_c$ measured by a Planck-brane observer is not affected by physics 
on the IR brane~\cite{Randall:1999vf,Goldberger:2002cz}.  For example, 
a momentum mode $p$ will only feel physics on the IR brane with a strength 
proportional to $e^{-\pi p/M_c}$.  Since $\Lambda_{\rm IR} > M_c$, any 
non-perturbative physics at $\Lambda_{\rm IR}$ in the Higgs sector will have 
decoupled from physics described by momenta $p > M_c$ on the Planck brane. 
This is the key feature of AdS space that allows IR physics to become 
non-perturbative without affecting physics on the UV brane above the 
scale $M_c$.  Any running of gauge couplings above $M_c$ will not feel 
any (possibly non-perturbative) physics on the IR brane. 

There remain two issues for model-building.  How can we transmit the electroweak 
symmetry breaking caused by the vacuum expectation values (VEVs) of $H^{(b)}_u$ 
and $H^{(b)}_d$ on the TeV brane to the quarks and leptons localized on the 
Planck brane?  And how can we maintain the MSSM prediction for gauge coupling 
unification?  Both of these issues are simultaneously resolved if we introduce two 
additional Higgs doublets $H^{(B)}_u$ and $H^{(B)}_d$ in the bulk.  These fields 
can interact both with the Planck and TeV branes and can transmit electroweak 
symmetry breaking.  Further, the addition of these fields is sufficient to preserve 
the MSSM prediction for gauge coupling unification, as we will see later.

\subsection{Model}

We now construct our model.  While most of our construction applies to 
general values of $\Lambda_{\rm IR}$, we concentrate on the case where 
$\Lambda_{\rm IR} \sim {\rm TeV}$, i.e. $kR \sim 10$, in what follows 
because it gives the largest upper bound on the Higgs-boson mass and allows 
a simple implementation of supersymmetry breaking.  We take the gauge 
group in the bulk to be $SU(5)$, which is broken by boundary conditions 
at the Planck brane~\cite{Goldberger:2002pc}.  Specifically, the 5D gauge 
supermultiplet, which can be decomposed into a 4D $N=1$ vector superfield 
$V$ and a 4D $N=1$ chiral superfield $\Sigma$, obeys the following set of 
boundary conditions:
\begin{equation}
  \pmatrix{V \cr \Sigma}(x^\mu,-y) 
  = \pmatrix{P V P^{-1} \cr -P \Sigma P^{-1}}(x^\mu,y), 
\qquad
  \pmatrix{V \cr \Sigma}(x^\mu,-y') 
  = \pmatrix{P' V P'^{-1} \cr -P' \Sigma P'^{-1}}(x^\mu,y'), 
\label{eq:bc-g}
\end{equation}
where $y' = y - \pi R$.  Here, $V$ and $\Sigma$ are both in the adjoint of 
$SU(5)$, and $P$ and $P'$ are $5 \times 5$ matrices acting on gauge space 
taken here as $P = {\rm diag}(+,+,+,-,-)$ and $P' = {\rm diag}(+,+,+,+,+)$. 
This reduces the gauge symmetry at low energies to be $SU(3)_C \times SU(2)_L 
\times U(1)_Y$ (321): only the 321 component of $V$ has a zero mode.  The 
characteristic scale for the KK tower is $M_c$, which is a factor of a few 
smaller than the IR cutoff scale $\Lambda_{\rm IR}$. 

We also introduce two bulk hypermultiplets $\{ H, H^c \}$ and $\{ \bar{H}, 
\bar{H}^c \}$ in the fundamental representation of $SU(5)$. Here, we have 
decomposed a hypermultiplet into two 4D $N=1$ chiral superfields, where 
$H({\bf 5})$, $H^c({\bf 5}^*)$, $\bar{H}({\bf 5}^*)$, $\bar{H}^c({\bf 5})$ 
are 4D chiral superfields with the numbers in parentheses representing 
their transformation properties under $SU(5)$.  The boundary conditions are 
given by
\begin{equation}
  \pmatrix{H \cr H^c}(x^\mu,-y) 
  = \pmatrix{-P H \cr P H^c}(x^\mu,y), 
\qquad
  \pmatrix{H \cr H^c}(x^\mu,-y') 
  = \pmatrix{P' H \cr -P' H^c}(x^\mu,y'), 
\label{eq:bc-h}
\end{equation}
for $\{ H, H^c \}$, and similarly for $\{ \bar{H}, \bar{H}^c \}$.  The zero 
modes then arise only from the $SU(2)_L$-doublet components of $H$ and 
$\bar{H}$, which we call $H^{(B)}_u \subset H$ and $H^{(B)}_d \subset \bar{H}$. 
In general, a bulk hypermultiplet $\{ \Phi, \Phi^c \}$ can have a mass term 
in the bulk, which is written as 
\begin{equation}
  S = \int\!d^4x \int_0^{\pi R}\!\!dy \, 
    \biggl[ e^{-3k|y|}\! \int\!d^2\theta\, c_\Phi k \Phi \Phi^c 
    + {\rm h.c.} \biggr],
\label{eq:bulk-mass}
\end{equation}
in the basis where the kinetic term is given by $S_{\rm kin} = \int\!d^4x 
\int\!dy\, [e^{-2k|y|} \int\!d^4\theta (\Phi^\dagger \Phi + \Phi^c 
\Phi^{c\dagger}) + \{ e^{-3k|y|} \int\!d^2\theta (\Phi^c \partial_y \Phi 
- \Phi \partial_y \Phi^c)/2 + {\rm h.c.} \}]$~\cite{Marti:2001iw}.  The 
parameter $c_\Phi$ controls the wavefunction profile of the zero mode --- 
for $c_\Phi > 1/2$ ($< 1/2$) the wavefunction of a zero mode arising from 
$\Phi$ is localized to the Planck (TeV) brane; for $c_\Phi = 1/2$ it is 
conformally flat.  For $H$ and $\bar{H}$ fields, we choose $c_H$ and 
$c_{\bar{H}}$ to be $1/2$ (or slightly larger than $1/2$):
\begin{equation}
  c_H \simeq c_{\bar{H}} \simeq \frac{1}{2},
\label{eq:Higgs-bulk-mass}
\end{equation}
so that the zero modes of $H^{(B)}_u$ and $H^{(B)}_d$ have (almost) 
conformally-flat wavefunctions.  These fields then play a role of transmitting 
electroweak symmetry breaking in the TeV-brane Higgs sector, which will be 
introduced later, to the matter sector ($H^{(B)}_u$ and $H^{(B)}_d$ are not 
completely identified with the Higgs doublets in the MSSM, as we will see 
below). The matter fields, $Q$, $U$, $D$, $L$ and $E$ for each generation, 
are introduced on the Planck brane with the Yukawa couplings to $H^{(B)}_u$ 
and $H^{(B)}_d$:
\begin{equation}
  S = \int\!d^4x \int_0^{\pi R}\!\!dy \,\, 
    2 \delta(y) \biggl[ \int\!d^2\theta \left( y_u Q U H^{(B)}_u 
    + y_d Q D H^{(B)}_d + y_e L E H^{(B)}_d \right) + {\rm h.c.} \biggr].
\label{eq:yukawa-321}
\end{equation}
(Alternatively, matter could be introduced in the bulk with the appropriate 
boundary conditions and wavefunctions localized towards the Planck brane by 
having $c \gg 1/2$. All our analyses apply to both the brane and bulk matter 
cases.)  With the above configuration of fields, the prediction for the 
low-energy gauge couplings are the same as in the MSSM at the leading order 
and proton decay rates are adequately suppressed~\cite{Goldberger:2002pc}. 
Small neutrino masses are also naturally obtained through the see-saw 
mechanism by introducing right-handed neutrinos $N$ on the Planck 
brane or in the bulk~\cite{Goldberger:2002pc}. 

We now introduce the Higgs sector on the TeV brane, which could become 
strongly coupled at the IR cutoff scale $\Lambda_{\rm IR} \sim {\rm TeV}$. 
We introduce three chiral superfields $S({\bf 1})$, $H'({\bf 5})$ and 
$\bar{H}'({\bf 5}^*)$ on the TeV brane with the numbers in parentheses 
representing the transformation properties under $SU(5)$ (the active gauge 
group on the TeV brane is $SU(5)$ so that any multiplet on this brane must 
be in a representation of $SU(5)$).  We introduce the superpotential of 
the following form on the brane:
\begin{equation}
  S = \int\!d^4x \int_0^{\pi R}\!\!dy \,\, 
    2 \delta(y-\pi R) \biggl[ e^{-3\pi kR}\! \int\!d^2\theta 
    \Bigl( \lambda S H' \bar{H}' + f(S) \Bigr) 
    + {\rm h.c.} \biggr],
\label{eq:NMSSM-TeV}
\end{equation}
where, as explained earlier, $f(S)$ is a function of $S$ with the general form 
given by $f(S) = - \Lambda_S^2 S - (M_S/2) S^2 - (\kappa/3) S^3 + \cdots$\,. 
Now, however, $\Lambda_S$ and $M_S$ are mass parameters of the order of $M_5$, 
or somewhat smaller, in the original 5D metric.  Since the sector living on the 
IR/TeV brane only needs to stay perturbative up to $\Lambda_{\rm IR}$, we only 
need to require $\lambda(\Lambda_{\rm IR}), \kappa(\Lambda_{\rm IR}) \simlt 2\pi$. 
This gives a large quartic coupling for the doublet components of $H'$ and 
$\bar{H}'$, which we call $H^{(b)}_u \subset H'$ and $H^{(b)}_d \subset \bar{H}'$ 
(these fields consist of parts of the MSSM Higgs doublets, as we will see 
shortly).  Once supersymmetry is broken, $H^{(b)}_u$ and $H^{(b)}_d$ (and $S$) 
will obtain VEVs, breaking the electroweak symmetry.  An important point is that, 
due to the properties of AdS space, the introduction of TeV-brane fields and/or 
superpotentials does not modify the physics at higher energies, including 
the evolution of the gauge couplings.  Therefore, the prediction for gauge 
coupling unification is still the same as that in the MSSM.

To complete the construction of the model, we have to connect the two sets 
of Higgs doublets, $H^{(B)}_u, H^{(B)}_d$ and $H^{(b)}_u, H^{(b)}_d$.  This 
can be done by introducing mixing terms between these fields on the TeV brane:
\begin{equation}
  S = \int\!d^4x \int_0^{\pi R}\!\!dy \,\, 
    2 \delta(y-\pi R) \biggl[ e^{-3\pi kR}\! \int\!d^2\theta 
    \left( M_{ud}^{1/2} H \bar{H}' + M_{du}^{1/2} \bar{H} H' \right) 
    + {\rm h.c.} \biggr],
\label{eq:mixing-TeV}
\end{equation}
where $M_{ud}$ and $M_{du}$ are parameters having the mass dimension of $1$ 
and of order $M_5$.  Note that the doublet components of $H$, $\bar{H}$, $H'$ and 
$\bar{H}'$ are $H^{(B)}_u$, $H^{(B)}_d$, $H^{(b)}_u$ and $H^{(b)}_d$, respectively. 
In the KK-decomposed 4D theory, interactions of Eqs.~(\ref{eq:NMSSM-TeV},~%
\ref{eq:mixing-TeV}) give the following supersymmetric masses for 
the Higgs doublets
\begin{equation}
  W = \pmatrix{ H^{(B)}_u \quad H^{(b)}_u \cr}
    \pmatrix{
      0 & e^{-\pi kR}\sqrt{\frac{M_{ud}}{\pi R}} \cr
      e^{-\pi kR}\sqrt{\frac{M_{du}}{\pi R}} & \lambda \langle S \rangle \cr}
    \pmatrix{ H^{(B)}_d \cr\cr H^{(b)}_d \cr}, 
\label{eq:4D-Higgs-mass}
\end{equation}
where (and below) $H^{(B)}_u$ and $H^{(B)}_d$ stand for the zero modes for 
these fields, and $S$ is canonically normalized in 4D.  Here, we have assumed 
that $M_{ud}$ and $M_{du}$ are of order $M_5$ or smaller and taken into 
account the volume suppression factor arising from wavefunctions of $H^{(B)}_u$ 
and $H^{(B)}_d$.  We have also left out the potential coupling of the singlet 
to the bulk Higgs fields which is more volume suppressed than the other terms 
and therefore negligible.  This gives the desired mixing between $H^{(B)}_u$ 
and $H^{(b)}_u$, and $H^{(B)}_d$ and $H^{(b)}_d$.  After supersymmetry is 
broken, a certain parameter region of the model leads to VEVs for one set of 
the Higgs doublets $H_u$ and $H_d$ parameterized by
\begin{eqnarray}
  H_u &=& \cos\theta_u H^{(b)}_u + \sin\theta_u H^{(B)}_u,
\label{eq:LE-Higgs-1}\\
  H_d &=& \cos\theta_d H^{(b)}_d + \sin\theta_d H^{(B)}_d,
\label{eq:LE-Higgs-2}
\end{eqnarray}
which we assume to be the case.  These fields $H_u$ and $H_d$, therefore, 
are the Higgs doublets responsible for electroweak symmetry breaking. 
Assuming that $S$ does not have a large supersymmetric mass, they 
have a large quartic term arising from the superpotential coupling 
$W = \lambda \cos\theta_u \cos\theta_d S H_u H_d$:
\begin{equation}
  V_H = \lambda^2 \cos^2\!\theta_u \cos^2\!\theta_d\, |H_u H_d|^2.
\label{eq:4D-quartic}
\end{equation}
Furthermore, these doublets also couple to the quarks and leptons through 
the Yukawa couplings of Eq.~(\ref{eq:yukawa-321}):
\begin{equation}
  W = y'_u \sin\theta_u\, Q U H_u + y'_d \sin\theta_d\, Q D H_d
    + y'_e \sin\theta_d\, L E H_d,
\label{eq:4D-Yukawa}
\end{equation}
giving the quark and lepton masses, where $y'_{u,d,e} \equiv y_{u,d,e}%
/(\pi R)^{1/2}$ are dimensionless coupling parameters.  Therefore, 
for reasonable $O(1)$ values for the mixing angles, interactions in 
Eqs.~(\ref{eq:4D-quartic},~\ref{eq:4D-Yukawa}) can give the required 
quark-lepton masses and a large quartic term for the Higgs fields. 

What if the IR scale $\Lambda_{\rm IR}$ is much larger than the TeV scale? 
In this case $M_{ud}$ and $M_{du}$ (and $\Lambda_S$ and $M_S$ in the function 
$f(S)$) must be chosen such that their values measured in terms of the 4D 
metric are of order TeV, i.e. $e^{-\pi kR} M_{ud} \sim e^{-\pi kR} M_{du} \sim 
{\rm TeV}$.  Alternatively, for the NMSSM, one could introduce an additional 
set of Higgs fields $H''$ and $\bar{H}''$ on the TeV brane with the superpotential 
term $\delta(y-\pi R) \{ \hat{M}_{ud}^{1/2} H''(a \bar{H} + b \bar{H}') 
+ \hat{M}_{du}^{1/2} \bar{H}''(c H + d H') \}$, in which case $\hat{M}_{ud}$ 
and $\hat{M}_{du}$ could be as large as $M_5$. 

\begin{figure}[t]
\begin{center}
  \includegraphics[scale=0.6]{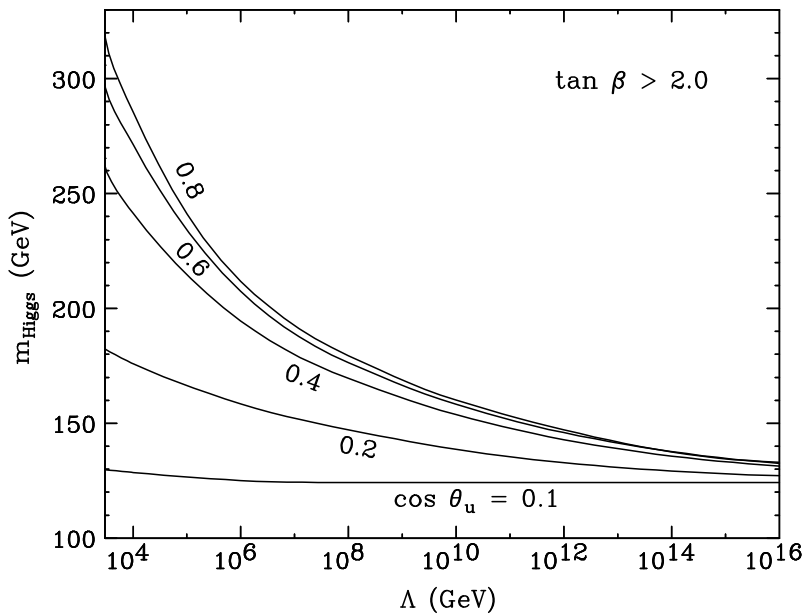}
\caption{Upper limits on the lightest Higgs-boson mass as a function of 
 the IR cutoff scale $\Lambda$ for various values of the up-type Higgs 
 mixing angle $\theta_u$: $\cos\theta_u = 0.1, 0.2, 0.4, 0.6$ and $0.8$.}
\label{fig:limplot-2}
\end{center}
\end{figure}
In Fig.~\ref{fig:limplot-2} we have plotted the upper limit on the physical 
Higgs-boson mass as a function of $\Lambda$.  In the figure, we have set the 
mixing angle for the down-type Higgs to be zero, $\theta_d = 0$, for simplicity, 
and plotted the limits for various different values of the mixing angle for 
the up-type Higgs, $\theta_u$.  The bound plotted is maximized by setting 
$\lambda(\Lambda) = 2\pi$ and $\kappa(\Lambda) = 0$, where $\Lambda \equiv 
\Lambda_{\rm IR}$, and by using the model with additional Higgs doublets $H''$ 
and $\bar{H}''$ described in the previous paragraph for higher values of 
$\Lambda$.  In the figure, we have applied the condition of $\tan\beta > 2$, 
because the values of $\tan\beta$ much smaller than $2$ require very small 
$\cos\theta_u$ to obtain a large enough top-quark mass and thus lead to a 
small physical Higgs-boson mass (the maximum Higgs-boson mass may actually be 
obtained at $\tan\beta$ somewhat smaller than 2, but the bound does not change 
much even in that case).  We have also assumed that the mixing of the two sets 
of Higgs doublets, Eqs.~(\ref{eq:LE-Higgs-1},~\ref{eq:LE-Higgs-2}), occurs at 
the scale $\Lambda$.  This maximizes a Higgs-boson mass obtained for a given 
value of $\Lambda$ and $\theta_u$.  In a realistic situation, however, the scale 
at which the Higgs mixing occurs will most likely be a factor of a few smaller 
than $\Lambda$.  This could give a reduction of the Higgs-boson mass as large 
as about $(30\!\sim\!50)~{\rm GeV}$ for $\Lambda \sim {\rm TeV}$ depending on 
details of the physics at $\sim \Lambda$, since the coupling $\lambda$ is then 
very large between $\Lambda$ and the scale of the Higgs mixing, which could 
give large negative corrections to the low-energy values of $\lambda$ and the 
top Yukawa coupling.  The reduction is smaller for larger values of $\Lambda$. 

The value of $\cos\theta_u$ is bounded from above by requiring that a large enough 
top-quark mass is obtained through the Yukawa coupling of Eq.~(\ref{eq:4D-Yukawa}). 
For example, if one neglects the effect of the running of $y_u$ (but not the 
volume-suppression effect), one finds that the upper bound on $\cos\theta_u$ 
is about $0.6$ for $\Lambda \simeq 10~{\rm TeV}$ and $\tan\beta \simgt 2$. 
The running effect between $\Lambda$ and the electroweak scale could somewhat 
reduce this value, e.g. to $\cos\theta_u \simlt 0.55$, but we still see from 
Fig.~\ref{fig:limplot-2} that our theory allows the Higgs-boson mass as high 
as $200~{\rm GeV}$, even after considering the potential reduction of the Higgs 
mass coming from physics at the scale $\Lambda$.  To complete the discussion, 
we must also consider the running of the top Yukawa coupling, $y_u$, above the 
scale $\Lambda$.  We will, however, see at the end of this section that the 
Higgs mass bound obtained here is not much changed by this effect.  Because 
our theory admits a large physical Higgs-boson mass, it potentially allows 
a significant reduction (or an elimination) of the fine-tuning.

Let us here consider an example of the parameter region leading to a large 
Higgs boson mass.  We here concentrate on the case with $\Lambda_{\rm IR} 
= O(10~{\rm TeV})$ and the region where the elements of the Higgs mass matrix 
in Eq.~(\ref{eq:4D-Higgs-mass}) are smaller than $M_c$.  In this parameter 
region mixings between the light Higgs states and higher KK states are 
negligible, so that we can neglect the effect of the KK states in analyzing 
electroweak symmetry breaking (note also that the precision electroweak 
constraints from the KK states are very weak with matter localized on 
the Planck brane~\cite{Davoudiasl:2000wi}).  For $M'_{du} \simlt M'_{ud} 
\simeq \lambda \langle S \rangle$, where $M'_{xy} \equiv e^{-\pi kR}
\sqrt{M_{xy}/\pi R}$ with $x,y=u,d$, we obtain $\theta_u = O(1)$ and 
$\theta_d \simlt 1$.  This leads to somewhat suppressed Yukawa couplings 
for the down-type quarks and charged leptons through Eq.~(\ref{eq:4D-Yukawa}), 
but not for the up-type quarks (especially the top quark).  The tree-level 
Higgs quartic couplings are given by Eq.~(\ref{eq:4D-quartic}) plus 
the contribution from the $SU(2)_L \times U(1)_Y$ $D$ terms.  The masses 
for two sets of Higgs doublets are of order $M_1 = M'_{ud} M'_{du}/\lambda 
\langle S \rangle$ and $M_2 = \lambda \langle S \rangle$, where $M_1 \simlt 
M_2$ (the mass of the Higgs triplets arising from $H'$ and $\bar{H}'$ is of 
order $M_2$).  Supersymmetry breaking is caused on the TeV brane by a 
non-zero VEV of a chiral superfield $Z$: $F_Z = \langle \partial_{\theta^2} 
Z \rangle \neq 0$, which gives masses for the gauginos through the operator 
$\delta(y-\pi R) \int\!d^2\theta Z {\cal W}^\alpha {\cal W}_\alpha$ and 
for the squarks and sleptons through finite loop contributions.%
\footnote{An alternative possibility is to break supersymmetry on the Planck 
brane with an intermediate scale VEV for $F_Z$, so that the 321 gauginos 
receive weak-scale masses on the Planck brane.  In this case, the Higgs 
fields $H'$ and $\bar{H}'$ do not obtain tree-level supersymmetry-breaking 
masses (and the supersymmetry-breaking masses for $H^{(B)}_u$ and 
$H^{(B)}_d$ are volume suppressed).  The squark and slepton masses, 
however, are generated at tree level through the couplings to $Z$, which 
must be assumed to be flavor universal.}
Then, if supersymmetry breaking masses for the Higgs fields, arising from 
operators of the form $\delta(y-\pi R) \sum_{i,j} \int\!d^4\theta Z^\dagger 
Z (H_i \bar{H}_j + H_i^\dagger H_j + \bar{H}_i^\dagger \bar{H}_j)$ where $H_i 
= H, H'$ and $\bar{H}_i = \bar{H}, \bar{H}'$, are of order $M_1$, correct 
electroweak symmetry breaking can be induced.  Supersymmetry breaking should 
also trigger a non-zero VEV for $S$, and we assume that the supersymmetric 
mass for $S$ -- e.g. $\kappa \langle S \rangle$ in the case with $M_S = 0$ 
-- is of order $M_1$.  Such a VEV, for example, may arise if the soft 
supersymmetry-breaking Lagrangian (and thus the superpotential) contains a 
linear term in $S$, ${\cal L}_{\rm soft} = (C_S \Lambda_S^2 S + {\rm h.c.}) 
+ \cdots$, in which case the VEV of $S$ is given by $\langle S \rangle 
\simeq C_S \Lambda_S^2/m_S^2$ in certain parameter region, where $m_S^2$ is 
the soft supersymmetry-breaking mass squared for $S$.  This allows relatively 
small values of $\langle S \rangle$ even with large values of $m_S^2$. 
The spectrum contains an extra pair of Higgs doublets and a pair of Higgs 
triplets with the masses of order $M_2$, in addition to the states in the 
NMSSM.  In fact, the presence of an extra pair of Higgs fields with the 
quantum numbers of ${\bf 5} + {\bf 5}^*$ under $SU(5)$ is a generic 
prediction of the model with $\Lambda_{\rm IR} \sim {\rm TeV}$. 

Finally, we consider the issue of evolution of the top Yukawa coupling above 
$\Lambda_{\rm IR}$: $y_u$ in Eq.~(\ref{eq:yukawa-321}) for the third generation. 
In the present model the evolution of the Yukawa coupling receives additional 
contribution from the bulk, which could potentially alter the existence and 
location of a Landau pole for the top Yukawa coupling (and for the other 
couplings).  While we do not make a full analysis of these coupling evolutions, 
we can make a rough estimate of this effect in the following way.  We first 
rescale the 5D Higgs field $H^{(B)}_u$ as $H^{(B)}_u \rightarrow \sqrt{M} 
\hat{H}_u$, so that $\hat{H}_u$ has a mass dimension of $1$.  Then the 
top Yukawa coupling of Eq.~(\ref{eq:yukawa-321}) is written as $2\delta(y) 
\int\!d^2\theta y_u \sqrt{M} Q U \hat{H}_u + {\rm h.c.}$ (note that $y_u$ 
has mass dimensions of $-1/2$).  Suppose that the fields $Q$, $U$ and 
$\hat{H}_u$ have brane-localized kinetic terms of $2\delta(y) \int\!d^4\theta 
(Z_{0,Q} Q^\dagger Q + Z_{0,U} U^\dagger U + Z_{0,H} \hat{H}^\dagger \hat{H})$ 
at tree level.  In the dual 4D picture, this implies that the wavefunction 
renormalizations for the fields $Q$, $U$ and $\hat{H}$, defined as 
$Z_Q(\mu)$, $Z_U(\mu)$ and $Z_H(\mu)$, take the values $Z_{0,Q}$, $Z_{0,U}$ 
and $Z_{0,H}$ at the scale $k$: 
\begin{equation}
  Z_Q(k)=Z_{0,Q},\qquad Z_U(k)=Z_{0,U}, \qquad Z_H(k)=Z_{0,H},
\end{equation}
where $\mu$ is the renormalization scale.  When we evolve the RGEs 
to low energies, $Z_Q$, $Z_U$ and $Z_H$ receive quantum corrections 
(but $y_u \sqrt{M}$ does not receive such corrections due to the 
non-renormalization theorem).  In particular, $Z_H(\mu=ke^{-\pi kR})$ 
receives a contribution from the bulk, which we interpret as the running 
effect: $\delta Z_{H,{\rm bulk}} = \pi R M = (M/k)\ln(k/\mu)$.  Then, if 
we simply add the bulk contribution to the MSSM running, we obtain RGEs 
for the wavefunctions: 
\begin{eqnarray}
  \frac{d\ln Z_Q}{d\ln\mu} &=& -\frac{1}{8\pi^2}
    \biggl( \frac{y_u^2 M}{Z_Q Z_U Z_H} - \frac{8}{3}g_3^2 \biggr),
\\
  \frac{d\ln Z_U}{d\ln\mu} &=& -\frac{1}{8\pi^2}
    \biggl( 2 \frac{y_u^2 M}{Z_Q Z_U Z_H} - \frac{8}{3}g_3^2 \biggr),
\\
  \frac{d\ln Z_H}{d\ln\mu} &=& -\frac{1}{8\pi^2}
    \biggl( 3 \frac{y_u^2 M}{Z_Q Z_U Z_H} \biggr) - \frac{M}{kZ_H}.
\end{eqnarray}
The low-energy top Yukawa coupling $y'_t$, which couples $Q$ and $U$ 
with the zero mode of $H^{(B)}_u$ (the 33 element of $y'_u$ in 
Eq.~(\ref{eq:4D-Yukawa})), is then obtained as 
\begin{equation}
  y'_t(\mu)^2 = \frac{y_u^2 M}{Z_Q(\mu) Z_U(\mu) Z_H(\mu)}. 
\end{equation}
Note that $Z_H(\mu)$ ($Z_Q(\mu)$ and $Z_U(\mu)$) is proportional to $M$ 
($M^0$), so that the coupling $y'_t$ does not depend on the spurious 
parameter $M$.  Using these RGEs, we can obtain the low-energy value of $y'_t$. 
Note that the $SU(3)_C$ gauge coupling $g_3$ obeys the RGE with the bulk 
contribution added. Assuming that the Planck-brane localized kinetic term 
at tree level is small, the RGE takes the form $d(1/g_3^2)/d\ln\mu = 
-(b/8\pi^2)$ with $b = b_{\rm MSSM} + b_{\rm bulk} \simeq 1.8$, where 
$b_{\rm bulk}$ represents the $SU(5)$-invariant bulk contribution, which 
makes $g_3$ non-perturbative at the scale $k$.  For natural sizes 
for the coefficients in 5D, i.e. $Z_{0,Q} \sim Z_{0,U} \sim 1$, $Z_{0,H} 
\sim M/M_*$ and $y_u \sim 4\pi/\sqrt{M_*}$, where $M_*$ is the 5D 
cutoff scale, we obtain $y'_t(\mu \sim {\rm TeV}) \simlt (1.3\!\sim\!1.4)$ for 
$\Lambda_{\rm IR} \sim {\rm TeV}$.  Although this estimate is somewhat 
uncertain, we expect that we can obtain a large enough top-quark mass 
$m_t = y'_t \sin\theta_u \langle H_u \rangle$, where $\langle H_u \rangle 
= v\, \sin\beta$, for $\sin\theta_u$ not much smaller than $1$, e.g. 
$\sin\theta_u \simgt 0.8$ for $\tan\beta \simgt 2$.  This in turn gives the upper 
limit on the lightest Higgs-boson mass larger than $200~{\rm GeV}$ for small 
values of $\Lambda_{\rm IR}$, since $\cos\theta_u$ as large as $0.6$ is 
allowed (see Fig~\ref{fig:limplot-2}).  Although a more careful analysis may 
be needed to be really conclusive about the issue of the top-quark mass, 
we expect that a Higgs boson mass as large as $200~{\rm GeV}$ can be 
obtained in this model.%
\footnote{It is possible to construct models in which the running of the 
top Yukawa coupling is the standard 4D running.  An example of such models 
is the following.  We have TeV-brane Higgs fields $S$, $H'$ and $\bar{H}'$ 
together with the superpotential of Eq.~(\ref{eq:NMSSM-TeV}).  We further 
introduce a Higgs field $H''_u$ on the Planck brane and a Higgs hypermultiplet 
$\{ \bar{H}, \bar{H}^c \}$ in the bulk, whose zero mode is denoted as 
$H^{(B)}_d$.  These fields are coupled to matter fields localized on the 
Planck brane as Eq.~(\ref{eq:yukawa-321}), but with $H^{(B)}_u$ replaced 
by $H''_u$.  Then, introducing the Higgs mixing terms of the form 
$\delta(y)\int\!d^2\theta H''_u H^{(B)}_d + \delta(y-\pi R)\int\!d^2\theta 
H^{(b)}_u H^{(B)}_d$, we obtain the top-Yukawa and Higgs-quartic couplings 
at low energies.  To get the 4D running for the top Yukawa coupling, 
however, we must arrange the coefficients of the two Higgs mixing terms 
such that they both have sizes around $\Lambda_{\rm IR}$ when measured 
in terms of the 4D metric $\eta_{\mu\nu}$.  Simple generalizations of this 
idea also lead to realistic down-type Yukawa couplings, while preserving 
the prediction for gauge coupling unification.}

\section{Model with Matter on the IR Brane}
\label{sec:model-2}

In this section we construct an alternative model.  This model differs from 
that of the previous section in the location of fields.  In particular, we 
locate the quark and lepton fields on the IR brane so that there is no issue 
of non-standard evolution of the Yukawa couplings (the evolution of the 
Yukawa couplings is the standard one in 4D below $\Lambda_{\rm IR}$). 
Because the Yukawa couplings are located on the IR brane, they can become 
non-perturbative at the scale $\Lambda_{\rm IR}$, giving the observed top-quark 
mass quite easily.  We again concentrate on the case with $\Lambda_{\rm IR} 
\sim {\rm TeV}$ below, since it gives the largest bound on the Higgs-boson 
mass and the simplest realization of gauge coupling unification.%
\footnote{The Higgs-boson mass bound in warped supersymmetric theories 
with matter fields on the TeV brane was also considered in~\cite{Casas:2001xv}. 
The model discussed there, however, does not accommodate the MSSM 
prediction for gauge coupling unification, and the bound on the Higgs mass 
is relaxed only for low values of $\Lambda$ of order TeV, as it uses operators 
suppressed by powers of $\Lambda$.}

The model uses a bulk $SU(5)$ gauge symmetry, but it is now broken 
at both the Planck and the TeV branes, i.e. the 5D gauge multiplet 
obeys the boundary conditions of Eq.~(\ref{eq:bc-g}) with $P = P' = 
{\rm diag}(+,+,+,-,-)$~\cite{Nomura:2004is}.  The bulk Higgs fields are 
introduced as two hypermultiplets in the fundamental representation of 
$SU(5)$, $\{ H, H^c \}$ and $\{ \bar{H}, \bar{H}^c \}$, with the boundary 
conditions given by Eq.~(\ref{eq:bc-h}) and $P = P' = {\rm diag}(+,+,+,-,-)$. 
These boundary conditions yield zero modes that are not the MSSM states, 
the $SU(5)/(SU(3)_C \times SU(2)_L \times U(1)_Y)$ component of $\Sigma$, 
in addition to the MSSM gauge fields, the 321 component of $V$.  However, 
these unwanted states can be made heavy (with masses of order TeV) through 
supersymmetry breaking effects on the TeV brane, and the prediction for 
the low-energy gauge couplings is still that of the MSSM, as long as 
the bulk mass parameters for the Higgs fields, $c_H$ and $c_{\bar{H}}$, 
satisfy $c_H, c_{\bar{H}} \geq 1/2$~\cite{Nomura:2004is}.  In this 
``321-321 model,'' the gauge groups effective at the Planck brane and 
the TeV brane are both $SU(3)_C \times SU(2)_L \times U(1)_Y$. Therefore, 
we can locate the MSSM matter fields $Q$, $U$, $D$, $L$ and $E$ on the 
TeV brane without introducing proton decay mediated by the $SU(5)$ gauge 
bosons.  Potentially dangerous tree-level proton decay operators can also 
be forbidden by imposing a symmetry, say baryon number, on the model.%
\footnote{We comment here that it is straightforward to use the 321-321 model 
to construct a theory of the type discussed in section~\ref{sec:model-1}, 
i.e. the model with the quarks and leptons on the Planck brane.  The simplest 
of such models has $\Lambda_{\rm IR} \sim {\rm TeV}$, with the boundary 
conditions for the gauge multiplet given by Eq.~(\ref{eq:bc-g}) with 
$P = P' = {\rm diag}(+,+,+,-,-)$ and those for the bulk Higgs hypermultiplets 
given by Eq.~(\ref{eq:bc-h}) with $P = -P' = {\rm diag}(+,+,+,-,-)$. 
The TeV-brane Higgs sector consists of $S$, $H^{(b)}_u$ and $H^{(b)}_d$ with 
the superpotential interactions of the form $\delta(y-\pi R) \int\!d^2\theta 
\{\lambda S H^{(b)}_u H^{(b)}_d + f(S)\}$.  The rest of the constructions 
of the model is analogous to those of section~\ref{sec:model-1}.}

We now take a closer look at the Higgs sector.  For the boundary conditions 
described above, the Higgs fields do not possess zero modes.  However, 
if the bulk mass parameters for the Higgs fields are much larger than 
$k/2$, i.e. $c_H, c_{\bar{H}} \gg 1/2$, four doublet states from $H$, 
$H^c$, $\bar{H}$ and $\bar{H}^c$ become exponentially lighter than $M_c$, 
which we assume to be the case (see footnote~2 of~\cite{Nomura:2004is}). 
Among these states, the modes arising from $H$ and $\bar{H}$, which we 
call $H'_u$ and $H'_d$ respectively, are (strongly) localized towards 
the Planck brane, while those arising from $H^c$ and $\bar{H}^c$, which 
we call $H_d$ and $H_u$ respectively, are (strongly) localized towards 
the TeV brane.  Now, we can introduce superpotential interactions between 
the $H_u$ and $H_d$ fields and the fields located on the TeV brane. 
Specifically, we introduce Yukawa couplings between the $H_u$ and $H_d$ 
fields and the quarks and leptons:
\begin{equation}
  S = \int\!d^4x \int_0^{\pi R}\!\!dy \,\, 
    2 \delta(y-\pi R) \biggl[ e^{-3\pi kR}\! \int\!d^2\theta 
    \left( y'_u Q U H_u + y'_d Q D H_d + y'_e L E H_d \right) 
    + {\rm h.c.} \biggr].
\label{eq:yukawa-TeV}
\end{equation}
In addition, we introduce a singlet chiral superfield $S$ on the TeV 
brane together with the superpotential couplings
\begin{equation}
  S = \int\!d^4x \int_0^{\pi R}\!\!dy \,\, 
    2 \delta(y-\pi R) \biggl[ e^{-3\pi kR}\! \int\!d^2\theta 
    \Bigl( \lambda' S H_u H_d + f(S) \Bigr) 
    + {\rm h.c.} \biggr],
\label{eq:singlet-TeV}
\end{equation}
where $f(S)$ is as before.  Since the light $H_u$ and $H_d$ fields 
are strongly localized towards the TeV brane, the couplings in 
Eqs.~(\ref{eq:yukawa-TeV},~\ref{eq:singlet-TeV}) do not receive a volume 
suppression factor when reduced to the low-energy 4D theory.  This allows 
large couplings for the low-energy superpotential 
\begin{equation}
  W = \lambda S H_u H_d + f(S) 
    + y_u Q U H_u + y_d Q D H_d + y_e L E H_d,
\label{eq:LE-W}
\end{equation}
where $y_{u,d,e} = y'_{u,d,e}\sqrt{k}$ and $\lambda = \lambda'\sqrt{k}$. 
Since $\lambda$ in Eq.~(\ref{eq:LE-W}) can be large of order $\pi$ at 
low energies of $E \simeq M_c$, a large physical Higgs-boson mass can 
be obtained. The upper limits on the Higgs-boson mass are essentially given 
by Fig.~\ref{fig:limplot} because there is no issue in obtaining a large 
enough top-quark mass.  Because of the bound on the masses of the KK 
states~\cite{Davoudiasl:2000wi}, we expect that $\Lambda_{\rm IR} \simgt 
100~{\rm TeV}$ in the present model.  However, this still allows a 
Higgs-boson mass as large as about $300~{\rm GeV}$. 

There are several issues in the present model.  First, small neutrino 
masses can be generated by introducing right-handed neutrino hypermultiplets 
$\{N, N^c\}$ in the bulk with the $(+,+)$ and $(-,-)$ boundary conditions 
for $N$ and $N^c$ respectively, and coupling them to the lepton doublets 
on the TeV brane through the operators of the form $\delta(y-\pi R) 
\int\!d^2\theta L N H_u$. Then, if the bulk masses for right-handed neutrinos 
are large, i.e. $c_N \gg 1/2$, we can naturally obtain small (Dirac) neutrino 
masses~\cite{Grossman:1999ra}.  Potentially large neutrino masses from the 
TeV-brane operators $\delta(y-\pi R) \int\!d^2\theta (L H_u)^2$ can be forbidden 
if we impose lepton number with charges $L(1), N(-1)$ on the model. Second, 
the $H'_u$ and $H'_d$ states, which are localized towards the Planck brane, 
must obtain a mass of order $\Lambda_{\rm IR}$ through the superpotential term 
$\delta(y) \int\!d^2\theta H'_u H'_d$ to evade phenomenological constraints 
and to preserve the MSSM prediction for gauge coupling unification.  The 
required mass term can naturally be generated by shining the scale from the 
TeV to Planck brane through a bulk singlet field with $c \simeq 1/2$, as 
discussed in~\cite{Goldberger:2002pc}.  Alternatively, for $\Lambda_{\rm IR} 
\sim {\rm TeV}$, the mass of the $H'_u$ and $H'_d$ states of order 
$\Lambda_{\rm IR}$ can be generated through supersymmetry breaking by 
introducing a singlet field $X$ on the Planck brane with the superpotential 
interaction of the form $\delta(y) \int\!d^2\theta (X H'_u H'_d + X^3)$, as in 
the NMSSM.  Finally, supersymmetry breaking on the TeV brane will give masses 
at tree level not only to the gauginos but also to the squarks and sleptons 
through the operators of the form $\delta(y-\pi R) \int\!d^4\theta Z^\dagger Z 
(Q^\dagger Q + U^\dagger U + D^\dagger D + L^\dagger L + E^\dagger E)$. 
This in general leads to the supersymmetric flavor problem, which we somehow 
have to avoid.  This may be accomplished, for example, by imposing a flavor 
symmetry such as $U(2)_F$ acting on the first two generations to these 
interactions.  Such a symmetry should be broken to generate realistic quark 
and lepton mass matrices, which in turn gives a small non-universality in 
the squark and slepton mass matrices.  Although it is not obvious that 
the supersymmetric flavor problem is naturally solved along this line, 
here we do not attempt to make detailed studies on this issue and simply 
assume that it can be done in a phenomenologically successful manner 
(if not, we have to make an assumption for the interactions between the 
$Z$ field and the quarks and leptons, which is not explained in our 
effective field theory). Since supersymmetry-breaking masses for the 
gauginos, sfermions and Higgs fields can naturally be of the same order, 
electroweak symmetry breaking can be obtained quite naturally in this model.

\section{Conclusions}
\label{sec:concl}

In the MSSM the upper bound on the mass of the lightest neutral Higgs 
boson is about $130~{\rm GeV}$.  We have constructed simple, realistic 
supersymmetric models which allow the mass of this particle to significantly 
exceed $130~{\rm GeV}$ while maintaining the MSSM prediction for gauge 
coupling unification.  These models are based on warped 5D versions of the 
NMSSM (or related theories).  The point is that by localizing the NMSSM 
singlet on the IR brane it is possible for the couplings of this field to 
become large at the IR cutoff without affecting the prediction of gauge 
coupling unification.  These can give a large contribution to the quartic 
coupling of the Higgs field, thereby raising the bound on the mass of 
the lightest neutral Higgs boson. 

Several versions of these models are possible, differing primarily 
in the location of the quark and lepton fields and in the pattern of 
breaking of the unified gauge symmetry.  The exact bound on the 
Higgs boson mass is somewhat different in each of these models. 
In this paper we have concentrated primarily on two particularly simple 
cases.  In the first, the quark and lepton fields are localized on the UV 
brane, where the unified symmetry is broken.  There are (at least) two 
sets of Higgs doublets --- one localized on the IR brane receiving 
a large quartic coupling from the NMSSM potential on the IR brane, 
the other propagating the bulk having the Yukawa couplings to the 
quarks and leptons.  The Higgs doublets responsible for electroweak 
symmetry breaking are linear combinations of these two sets, thus 
having both the Yukawa couplings and a large quartic coupling. 
The running of the top Yukawa coupling, however, is non-standard 
above the IR scale and it limits the maximum value of the bound we 
can obtain in this model.  In the second model, the quark and lepton 
fields are localized on the IR brane, while the unified symmetry is 
broken both on the UV and IR branes.  In this case there is no issue 
of the non-standard running of the top Yukawa coupling so that we 
can obtain the maximal value of the bound on the Higgs mass, which 
can be as large as $300~{\rm GeV}$.

Through the AdS/CFT correspondence, our theory is related to purely 
4D theories in which some (or all) of the Higgs fields are composites 
of some strong interaction, which is nearly conformal above the scale 
of the IR cutoff. In this sense our theory shares certain features with 
the model considered in~\cite{Harnik:2003rs}.  In our model, 
however, the prediction of gauge coupling unification is ``automatically'' 
the same as the MSSM, which is not the case in the model 
of~\cite{Harnik:2003rs}.  The issues of raising the Higgs mass bound 
and the prediction of the gauge couplings are also considered 
in~\cite{Chang:2004db}.  In that model, however, in contrast to ours, 
the Higgs doublets are mainly elementary while the singlet is composite.

\section*{Acknowledgments}

Z.C. would like to thank Nima Arkani-Hamed and Spencer Chang for discussions. 
A.B. would like to thank Kaustubh Agashe for helpful discussions and also 
Lawrence Berkeley National Laboratory and Cornell University for their kind 
hospitality during completion of this work.  This work was supported in 
part by the Director, Office of Science, Office of High Energy and Nuclear 
Physics, of the US Department of Energy under Contract DE-AC03-76SF00098 
and DE-FG03-91ER-40676 and in part by the National Science Foundation under 
grant PHY-0098840 and PHY-0403380.  The work of Y.N. was also supported by 
a DOE Outstanding Junior Investigator award.

\end{document}